\documentclass[final]{aipproc}
\layoutstyle{6s}

\title{}

\begin{document}

\title{The GRB~980425-SN1998bw Association in the EMBH Model}

\author{F. Fraschetti}{address={Universit\`a di Trento, Via Sommarive 14, I-38050 Povo (Trento), Italy}}

\author{M.G. Bernardini}{address={ICRA - International Centre for Relativistic Astrophysics and Dipartimento di Fisica, Universit\`a di Roma ``La Sapienza'', Piazzale Aldo Moro 5, I-00185 Roma, Italy}}

\author{C.L. Bianco}{address={ICRA - International Centre for Relativistic Astrophysics and Dipartimento di Fisica, Universit\`a di Roma ``La Sapienza'', Piazzale Aldo Moro 5, I-00185 Roma, Italy}}

\author{P. Chardonnet}{address={Universit\'e de Savoie, LAPTH - LAPP, BP 110, F-74941 Annecy-le-Vieux Cedex, France}}

\author{R. Ruffini}{address={ICRA - International Centre for Relativistic Astrophysics and Dipartimento di Fisica, Universit\`a di Roma ``La Sapienza'', Piazzale Aldo Moro 5, I-00185 Roma, Italy}}

\author{S.-S. Xue}{address={ICRA - International Centre for Relativistic Astrophysics and Dipartimento di Fisica, Universit\`a di Roma ``La Sapienza'', Piazzale Aldo Moro 5, I-00185 Roma, Italy}}

\begin{abstract}
Our GRB theory, previously developed using GRB~991216 as a prototype, is here applied to GRB~980425. We fit the luminosity observed in the 40--700 keV, 2--26 keV and 2--10 keV bands by the BeppoSAX satellite. In addition the supernova SN1998bw is the outcome of an ``induced gravitational collapse'' triggered by GRB~980425, in agreement with the GRB-Supernova Time Sequence (GSTS) paradigm (\citet{lett3}). A further outcome of this astrophysically exceptional sequence of events is the formation of a young neutron star generated by the SN1998bw event (\citet{cospar02}). A coordinated observational activity is recommended to further enlighten the underlying scenario of this most unique astrophysical system.
\end{abstract}

\maketitle

Our GRB theory (\citet{lett1,lett2,lett3,rbcfx02_letter,Brasile} and references therein), previously successfully applied to GRB~991216 used as a prototype, is applied to GRB~980425 (\citet{pian00}) and SN1998bw (\citet{g98}). This event allows to test the validity of the theory over a range of energies of 6 orders of magnitude: both sources appear to be spherically symmetric and the respective total energies are $E_{tot}\simeq 5\times 10^{53}\, {\rm ergs}$ and $E_{tot}\simeq 10^{48}\, {\rm ergs}$. 

The theory, therefore, explains all the observed features of the bolometric intensity variations of the afterglow as well as the spectral properties of the source and, in the specific case of GRB~980425 (\citet{cospar02}), it also allows to clarify the general astrophysical scenario in which the GRB actually occurs. In this system, in fact, we propose that GRB~980425 has been the trigger of a phenomenon of ``induced gravitational collapse'' (\citet{lett3}) originating the supernova explosion and we also witness the birth of a young neutron star out of the supernova event. This extraordinary coincidence of these three astrophysical events represents an unprecedented scenario of fundamental importance in the field of relativistic astrophysics. 

The observational situation of this system is quite complex. In addition to the source GRB~980425 and the supernova SN1998bw, two X-ray sources have been found by BeppoSAX in the error box for the location of GRB~980425: a source {\em S1} and a source {\em S2} (\citet{pian00}). Our approach is the following. We first interpret the GRB~980425 within the EMBH theory. This allows the computation of the luminosity, spectra, Lorentz gamma factors, and more generally all the dynamical aspects of the source. Having characterized the features of GRB~980425, we can gradually approach the remaining part of the scenario, disentangling the GRB observations from the supernova ones and from the sources {\em S1} and {\em S2}. This leads to a natural time sequence of events and to their autonomous astrophysical characterization.

Our approach has focused on identifying the energy extraction process from the black hole (\citet{cr71}) as the basic energy source for the GRB phenomenon. The distinguishing feature is a theoretically predicted source energetics all the way up to $1.8\times 10^{54}\left(M_{BH}/M_{\odot}\right) {\rm ergs}$ for $3.2 M_{\odot} \le M_{BH} \le 7.2 \times 10^6 M_{\odot}$ (\citet{dr75}). In particular, the formation of a ``dyadosphere'', during the gravitational collapse leading to a black hole endowed with electromagnetic structure (EMBH) has been indicated as the initial boundary conditions of the GRB process (\citet{r98,prx98}). 

The equations of motion in our theory depend only on two free parameters: the total energy $E_{tot}$, which coincides with the dyadosphere energy $E_{dya}$, and the amount $M_B$ of baryonic matter left over from the gravitational collapse of the progenitor star, which is determined by the dimensionless parameter $B=M_Bc^2/E_{dya}$. Our best fit corresponds to $E_{dya}=1.1\times 10^{48}\, {\rm ergs}$, $B=7\times 10^{-3}$ and the ISM average density is found to be $\left<n_{ism}\right>=0.02~{\rm particle}/{\rm cm}^3$. The plasma temperature and the total number of pairs in the dyadosphere are respectively $T=1.028\, {\rm MeV}$ and $N_{e^\pm}=5.3274\times10^{53}$.

Recently, within the EMBH theory, we have developed an attempt to theoretically derive the GRB spectra out of first principles as well as the GRB luminosity in fixed energy bands (\citet{Spectr1}). We have adopted three basic assumptions: {\bf a)} the resulting radiation as viewed in the comoving frame during the afterglow phase has a thermal spectrum and {\bf b)} the ISM swept up by the front of the shock wave, with a Lorentz gamma factor between $300$ and $2$, is responsible for this thermal emission. {\bf c)} We also assume, like in our previous papers (\citet{lett1,lett2,rbcfx02_letter,Brasile}), that the expansion occurs with spherical symmetry. 

The temperature $T$ of the black body in the comoving frame is then
\begin{equation}
T=\left(\frac{\Delta E_{\rm int}}{4\pi r^2 \Delta \tau \sigma {\cal R}}\right)^{1/4}\, ,
\label{tcom}
\end{equation}
where ${\cal R}={A_{eff}}/{A_{abm}}$ is the ratio between the ``effective emitting area'' and the ABM pulse surface $A_{abm}$ (in this case the best fit value of ${\cal R}$ is monotonically decreasing from $4.81\times 10^{-10}$ to $2.65\times 10^{-12}$), $\sigma$ is the Stefan-Boltzmann constant and $\Delta E_{\rm int}$ is the proper internal energy developed in the collision between the ABM pulse and the ISM in the proper time interval $\Delta \tau$ (see \citet{Brasile,Spectr1}). The ratio ${\cal R}$, which is a priori a function that varies as the system evolves, is evaluated at every given value of the laboratory time $t$.

All the subsequent steps are now uniquely determined by the equations of motion of the system. The basic tool in this calculation involves the definition of the EQuiTemporal Surfaces (EQTS) for the relativistic expanding ABM pulse as seen by an asymptotic observer. The key to determining such EQTS (see Fig.~1 in \citet{rbcfx02_letter}) is the relation between the time $t$ in the laboratory frame at which a photon is emitted from the ABM pulse external surface and the arrival time $t_a^d$ at which it reaches the detector. 

The results are given in Fig.~1 where the luminosity is computed as a function of the arrival time for three selected energy bands.

In Fig.~1 the luminosities in the three bands are represented together with the optical data of SN1998bw (black dots), the source S1 (black squares) and the source S2 (open circles). It is then clear that GRB~980425 is separated both from the supernova data and from the sources {\em S1} and {\em S2}.

\begin{figure}
\includegraphics[width=\hsize]{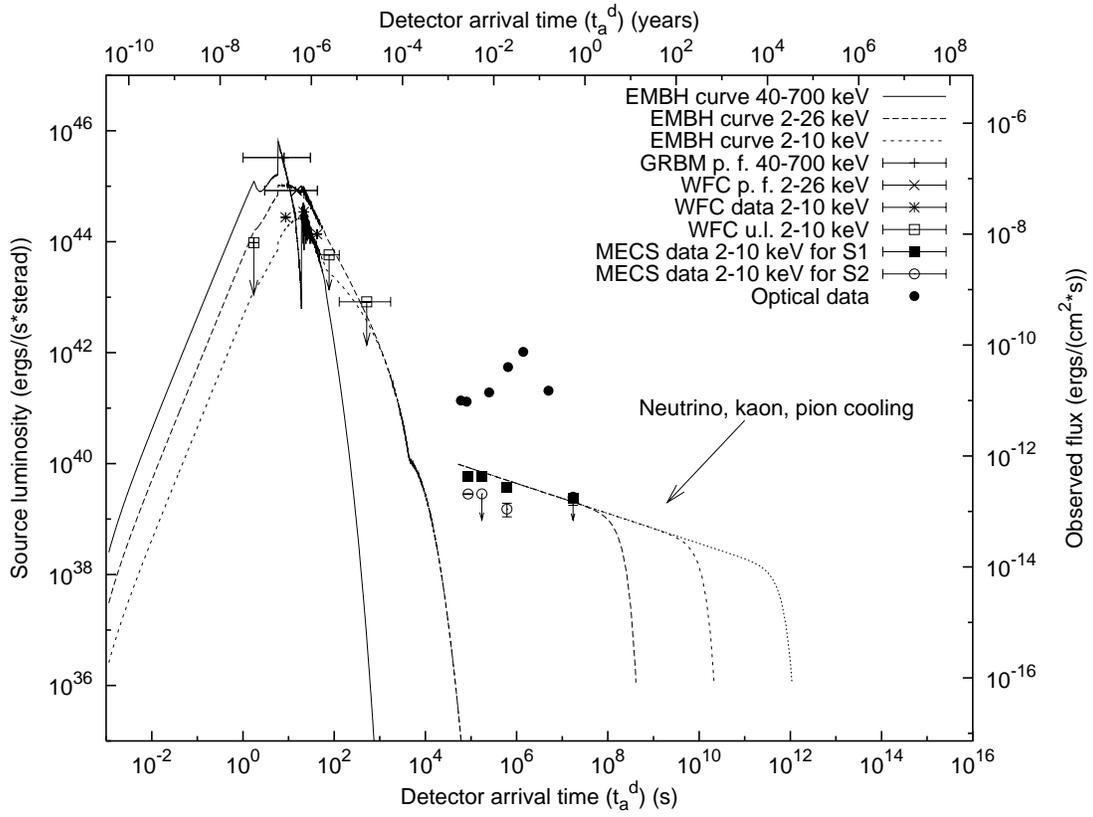}
\caption{The light curves in selected bands are reported as well as the MECS light curves in the 2-10 keV band of S1 and S2 (\citet{pian00}) as well as the optical data (\citet{i99}). Here are also reported theoretical models of neutron star cooling (\citet{c78}).}
\end{figure}

While the occurrence of the supernova in relation to the GRB has already been discussed with the GRB-Supernova Time Sequence (GSTS) paradigm (\citet{lett3}), we like to address here a different fundamental issue: the possibility of observing the birth of a newly formed neutron star, possibly pulsating, out of the supernova event, which in turn has been triggered by the GRB~980425.

In the early days of neutron star physics it was clearly shown by (\citet{gs41}) that the URCA processes are at the very heart of the supernova explosions. The neutrino-antineutrino emission described in the URCA process is the essential cooling mechanism necessary for the occurrence of the process of gravitational collapse of the imploding core. Since then, it has become clear that the newly formed neutron star can be still significantly hot and in its early stages will be associated to three major radiating processes (\citet{t64,t79,t02,c78}): {\bf a)} the thermal radiation from the surface, {\bf b)} the radiation due to neutrino, kaon, pion cooling, and {\bf c)} the possible influence in both these processes of the superfluid nature of the supra-nuclear density neutron gas. Qualitative representative curves for these cooling processes, which are still today very undetermined due to the lack of observational data, are shown in Fig.~1.

It is of paramount importance to follow the further time history of the two sources {\em S1} and {\em S2}. If, as we propose, {\em S2} is a background source, its flux should be practically constant in time and this source has nothing to do with the GRB~980425~/~SN1998bw system. If S1 is indeed the cooling radiation emitted by the newly born neutron star, it should be possible to notice a very drastic behavior in its luminosity as qualitatively expresses in Fig.~1.

The complete details on the source with all numerical values and explicit relations is going to appear in (\citet{rbbcfx04}).

\bibliographystyle{aipproc}

\bibliography{LosAlamos_2003}

\end{document}